\documentclass[%
 reprint,
 amsmath,amssymb,
 aps,
 prd,
]{revtex4-2}

\usepackage{amsmath,amsfonts,amssymb,amsthm,amstext,amscd,eucal,xcolor}
\usepackage{color}
\usepackage[all]{xy}
\usepackage{hyperref}
\usepackage{epsfig}
\usepackage[active]{srcltx}

\usepackage{graphicx}
\usepackage{dcolumn}
\usepackage{bm}

\newcommand{\be}{\begin{equation}}
\newcommand{\ee}{\end{equation}}
\newcommand{\bea}{\begin{eqnarray}}
\newcommand{\eea}{\end{eqnarray}}
\newcommand{\bse}{\begin{subequations}}
\newcommand{\ese}{\end{subequations}}
\newcommand{\beqa}{\begin{eqnarray}}
\newcommand{\eeqa}{\end{eqnarray}}
\newcommand{\beqar}{\begin{eqnarray*}}
\newcommand{\eeqar}{\end{eqnarray*}}
\newcommand{\bi}{\begin{itemize}}
\newcommand{\ei}{\end{itemize}}
\newcommand{\bn}{\begin{enumerate}}
\newcommand{\en}{\end{enumerate}}

\newcommand{\ba}{\begin{array}}
\newcommand{\ea}{\end{array}}
\newcommand{\bc}{\begin{center}}
\newcommand{\ec}{\end{center}}

\def\a{\alpha}

\definecolor{darkgreen}{rgb}{0,0.3,0}
\definecolor{mgreen}{rgb}{0,0.6,0}
\definecolor{darkblue}{rgb}{0,0,0.3}
\definecolor{darkred}{rgb}{0.7,0,0}

\newcommand{\pdiff}{\partial}

\def\be{\begin{equation}}
\def\ee{\end{equation}}

\begin{document}

\preprint{APS/123-QED}

\title{Separability of Klein-Gordon Equation on Near Horizon Extremal \\ Myers-Perry Black Hole}

\author{Hovhannes Demirchian}
\email{demhov@gmail.com}
 \affiliation{\it Yerevan Physics Institute, 2 Alikhanian Brothers St., 0036, Yerevan, Armenia.}


\author{Saeedeh Sadeghian}%
 \email{s.sadeghian@umz.ac.ir}
\affiliation{
 \it Department of Theoretical Physics, Faculty of Basic Sciences,
University of Mazandaran, P.O. Box 47416-95447, Babolsar, Iran
}%
\affiliation{\it  
ICRANet-Mazandaran, University of Mazandaran, P. O. Box 47415-416, Babolsar, Iran.
}%


\begin{abstract}
\noindent
We investigate the separability of Klein-Gordon equation on near horizon of d-dimensional rotating Myers-Perry black hole in two limits : 1) generic extremal case and 2) extremal vanishing horizon case. In the first case , there is a relation between the mass and rotation parameters so that black hole temperature vanishes. In the latter case, one of the rotation parameters is restricted to zero on top of the extremality condition. We show that the Klein-Gordon equation is separable in both cases. Also, we solved the radial part of that equation and discuss its behaviour in small and large r regions. 
\end{abstract}
\setcounter{footnote}{0}

\maketitle

\section{Introduction}\label{sec-intro}

The four dimensional Kerr black hole has been widely studied from various aspects. From the geometric point of view, Carter showed that it has integrable geodesics\cite{Carter:1968rr}. 
The extension to higher dimensions is known as Myers-Perry black hole. In d-dimensions, it is described by one mass parameter and $N(=[\frac{d}{2}])$ number of rotation parameters. It has integrable geodesic equations just like its 4-dimensional equivalent. Other extensions such as including the cosmological constant and NUT charge does not change this behavior\cite{Frolov:2006pe,Krtous2007,PhysRevLett.98.061102}. The integrability of the Klein-Gordon and Maxwell field equations have been also studied on the d-dimensional Kerr-(A)dS-NUT space-time\cite{Frolov:2006pe,Krtous2007,Cariglia2011,Kolar:2015hea,Lunin2017,Krtous2018}. 
 
One can construct another solution to Einstein equations in the Near Horizon extremal limit \cite{PhysRevD.60.104030,Kunduri2007,Kunduri2009,Kunduri2013} of MP (NHEMP) black hole \cite{Figueras2008} (see \cite{Hakobyan:2017qee,Demirchian2018,Galajinsky2013,Galajinsky2013a,Demirchian:2017uvo, Kolar:2017vjl} for recent studies). The extremal limit of the parameters describes a black hole with the biggest allowed angular momentum for a given BH mass. When one of the rotation parameters of the BH vanishes, we arrive to yet another solution of Einstein equations referred to as extremal vanishing horizon (EVH) geometry\cite{DeBoer2012,Sheikh-Jabbari2011}. Although the generic extremal limit of Myers-Perry black hole exists in both even and odd dimensions, the special EVH limit exists only in odd-dimensions.

A set of the Killing vectors of NHEMP obey the structural relation of $SL$(2,R) algebra corresponding to  $AdS_2$ subspace.  It has been demonstrated (e.g. \cite{Hakobyan2009,Hakobyan2010, Galajinsky2011, Galajinsky2008}) that the Casimir element of this $SO$(2, 1) algebra gives rise to a reduced Hamiltonian system called spherical or angular mechanics, which contains all the necessary information about the near horizon geometry. In other words, a massive particle moving in the near horizon geometry of an extremal rotating black hole possesses a dynamical conformal symmetry, i.e.  defines ``conformal mechanics" \cite{Galajinsky2013a,Galajinsky2013,Hakobyan:2017qee,Demirchian:2017uvo, Galajinsky2008,Galajinsky2010,Galajinsky2011,Galajinsky2012,Galajinsky2011a,Bellucci2011,Saghatelian2012,Galajinsky2016}, whose Casimir element can be viewed as a reduced Hamiltonian, which contains all the necessary information about the whole system.

An eye-catching difference between NHEMP and its vanishing horizon limit is that the latter has a larger isometry group\cite{SADEGHIAN2015222,SADEGHIAN2016488,PhysRevD.96.044004}: it includes two copies of SL(2,R) corresponding to the $AdS_3$ subspace instead of one copy of SL(2,R) for the $AdS_2$ factor of the metric. This symmetry enhancement does not add to number of independent constants of motion though \cite{Demirchian2018}.

The near horizon geometry of Myers-Perry black holes contains integrable and superintegrable systems like Rosochatius and P\"{o}schl-Teller which are interestingly related to Klein-Gordon equation through a geometrization procedure \cite{Evnin2017}. The angular part of near horizon limit of fully isotropic Myers-Perry black hole is a superintegrable mechanics called Rosochatius system. This is a direct generalization of the Higgs oscillator. Separation of variables in Rosochatius system results into a recursive family of one-dimensional P\"{o}schl-Teller system. The quantum equivalents of Higgs oscillator, Rosochatius and P\"{o}schl-Teller systems can be associated with a Klein-Gordon equation on a static spacetime.

In this work, we study the separability of Klein-Gordon equation on near horizon geometries of odd-dimensional Myers-Perry black hole. First, we take the near horizon geometry in the generic extremal case and use the elliptical coordinates in which the geodesic equation is separable. The SL(2,R)$\times$U(1)$^N$ isometry group of the background metric helps us to simply separate the part of Klein-Gorodon equation related to $AdS_2$ subspace and azimuthal angles. We observe that the Klein-Gordon equation is separable in the elliptical coordinates. We also find the solution to the radial part of the Klein-Gordon equation and discuss some general properties of the solution. Then, we study the near horizon EVH space-time and show the separability of the Klein-Gordon equation on that metric.

		
\section{Klein-Gordon equation on near horizon extremal geometry}
	
The near horizon extremal metric of odd dimensional $(d=2N+1)$ Myers-Perry black hole (NHEMP) in Gaussian null coordinates was given in \cite{Figueras2008} and can be written in the Boyer-Lindquist coordinates as
\bea
ds^2=&&\frac{F_H}{b(r_H)}\left(-r^2dt^2+\frac{dr^2}{r^2}\right)\nonumber\\
&&+\sum_{i=1}^{N}(r_H^2+a_i^2)d\mu_i^2+\sum_{i,j=1}^N \gamma_{ij}D\varphi^i D\varphi^j\,,
\eea
where,
\be
D\varphi^i\equiv d\varphi^i+\frac{B^i}{b} rd\tau, \qquad B^i=\frac{2}{r_H^2}\frac{\sqrt{m_i-1}}{m_i^2}\,.
\ee

The metric functions are
\be
\begin{gathered}
	F_H=\sum_{i=1}^{N}\frac{\mu_i^2}{m_i}, 
	\qquad b(r_H)=\frac{4}{r_H^2}\sum_{i<j}^{N}\frac{1}{m_i}\frac{1}{m_j}\,,\\
	\gamma_{ij}=(r_H^2+a_i^2)\mu_i^2\delta_{ij}+\frac{1}{F_H}a_i\mu_i^2a_j\mu_j^2\,,
\end{gathered}
\ee
Here, $m_i$'s are some constant parameters related to the horizon radius ($r_H$) and $N$ number of rotation parameters ($a_i$'s) corresponding to azimuthal coordinates $\varphi^i$ defined by,
\be
m_i=\frac{r_H^2+a_i^2}{r_H^2} \ge 1\,,
\ee
and $\mu_i$ are the latitudinal coordinates which satisfy the following relation
\be
\label{eq:mu_sum}
\sum_{i=1}^{N}\mu_i^2=1\,.
\ee
The location of the horizon $r_H$ is determined by the largest positive solution of 
\be
\label{eq:r_H}
\Pi(r_H)=m\, r_H^2,
\ee
where $\Pi(r)\equiv\prod_{i=1}^{N}(r^2+a_i^2)$ and  $m$ is a constant related to the mass of the Myers-Perry black hole. 

The extremal limit is given by
\be
\label{eq:extral_limit}
\Pi'(r_H)=2 m \, r_H\,.
\ee
By combining these two relations one finds that $m_i$'s are restricted by

\be
\label{eq:parameters_odd}
	\sum_{i=1}^{N}\frac{1}{m_i}=1\,.
\ee
By substituting $\mu_i=x_i / \sqrt{m_i}$ for $i=1\ ...\ N$, the new form of NHEMP metric in an arbitrary odd dimensions becomes
\bea
\label{eq:nhemp_x}
	\frac{ds^2}{r^2_H}=&&A(x)\left(-r^2d\tau^2+\frac{dr^2}{r^2}\right)\nonumber\\
	&&+\sum_{i=1}^N dx_idx_i+
	\sum_{i,j=1}^N\tilde{\gamma}_{ij}x_i x_j D\varphi^iD\varphi^j\,,
\eea
where 
\bea
\label{m2_odd}
&&D\varphi^i\equiv d\varphi^i+k^ird\tau,\qquad k^i=\frac{B^i}{b}\,,\nonumber\\
&&A(x) =\frac{\sum_{i=1}^N x^2_i/m^2_i}{r_H^2\ b}, \qquad \sum_{i=1}^N \frac{x^2_i}{m_i}=1\,,\nonumber\\ 
&&\tilde{\gamma}_{ij}=\delta_{ij}+ \frac{1}{\sum_l^N x_l^2/m^2_l}\frac{\sqrt{m_i-1}x_i}{m_i}  \frac{\sqrt{m_j-1}x_j}{m_j}\,.
\eea


The behavior of a massive scalar field $\Phi$ in the gravitational background is
governed by the Klein-Gordon equation:
\be
	\label{eq:KG}
	\square\Phi=\frac{1}{\sqrt{-g}}{\pdiff_\alpha(\sqrt{-g}g^{\alpha\beta}\pdiff_\beta\Phi)}=M^2\Phi\,,
\ee
where $M$ is the mass of the scalar field and $g$ is the determinant of the metric. Separation of variables of Klein-Gordon equation in the background of odd dimensional NHEMP can be carried out in elliptic coordinates $\lambda_a$ which is related to $x_i$ with the following relation
\be
	\label{xN}
	x^2_i=(m_i-\lambda_i)\prod_{j=1, j\neq i}^{N}\frac{m_i-\lambda_j}{m_i-m_j}\,,
\ee
where $\lambda_N  < m_N  <  \ldots < \lambda_2  < m_2  < \lambda_1 < m_1$. To resolve the last relation in \eqref{m2_odd} one should choose $\lambda_N=0$. In these  coordinates $\lambda_a$
\be
	\label{eq:h}
	\sum_{i=1}^N dx_i^2=\sum_{a=1}^{N-1} h_a(\lambda)d\lambda_{a}^{2}\,,
\ee
with
\be
h_a=-\frac{\lambda_{a}\prod_{b\neq a}^{N-1}(\lambda_b-\lambda_a)}{4\prod_{i=1}^{N} (m_i-\lambda_a)}\,.
\ee
the NHEMP metric \eqref{eq:nhemp_x} becomes
\bea
	\label{eq:nhemp}
	\frac{ds^2}{r^2_H}=&&A(\lambda)\left(-r^2d\tau^2+\frac{dr^2}{r^2}\right)+\sum_{a=1}^{N-1} h_a(\lambda)d\lambda_{a}^{2}\nonumber\\
	&&+\sum_{i,j=1}^N\tilde{\gamma}_{ij}x_i(\lambda) x_j(\lambda) D\varphi^iD\varphi^j,
\eea
here 
\be
	A(\lambda)=\frac{1}{b}\frac{\prod\limits_{a=1}^{N-1}\lambda_a}{\prod\limits_{i=1}^{N}m_i}.
\ee
To analysis the Klein-Gordon equation \eqref{eq:KG}, we need to find the determinant and the inverse metric of \eqref{eq:nhemp}. 
The inverse of NHEMP metric is
\be
\label{eq:nhemp_inv}
\begin{aligned}
	r^{2}_H\left(\frac{\partial}{\partial s}\right)^2=&-\frac{1}{A(\lambda)r^2}\left(\frac{\pdiff}{\pdiff\tau}-\sum_{i=1}^{N}r\ k^i\frac{\pdiff}{\pdiff\varphi^i}\right)^2\nonumber\\
	&+\frac{r^2}{A(\lambda)}\left(\frac{\pdiff}{\pdiff r}\right)^2+\sum_{a=1}^{N-1} h^a(\lambda)\left(\frac{\pdiff}{\pdiff \lambda_{a}}\right)^{2}\nonumber\\
	&+\sum_{i,j=1}^N\tilde{\gamma}^{ij}\frac{1}{x_i(\lambda)} \frac{1}{x_j(\lambda)} \frac{\pdiff}{\pdiff \varphi^i}\frac{\pdiff}{\pdiff \varphi^j}\,,
\end{aligned}
\ee
where $h^a(\lambda)$ and $\tilde{\gamma}^{ij}$ are the inverses of $h_a(\lambda)$ and $\tilde{\gamma}_{ij}$ respectively
\be
	h^a(\lambda)=h_a^{-1}(\lambda), \qquad \tilde{\gamma}^{ij}=\delta^{ij}-x_i\frac{\sqrt{m_i-1}}{m_i}x_j\frac{\sqrt{m_j-1}}{m_j}.
\ee

The determinant of metric \eqref{eq:nhemp} has the following form
\be
	\label{eq:det_g_initial}
	-\det{g}=A(\lambda)r^2 \cdot \frac{A(\lambda)}{r^2} \cdot \prod_{a=1}^{N-1} h_a \cdot  det\left(\tilde{\gamma}_{ij}x_i x_j\right).
\ee
Taking into account the definition of $h_a$ \eqref{eq:h} and the relation \eqref{xN}, $\prod_{a}h_a$ simplifies to
\bea
	\label{eq:prod_h}
	&&\prod_{a=1}^{N-1} h_a=c\cdot	\frac{A(\lambda)\left(\prod\limits_{a<b}^{N-1}(\lambda_b-\lambda_{a})\right)^2}{\prod\limits_{i=1}^{N}x_i^2}\nonumber\\, 
	&&c=\frac{(-1)^{\frac{N(N-1)}{2}}b\left(\prod\limits_{i=1}^{N}m_i\right)^2}{4^{N-1}\prod\limits_{\substack{i,j=1 \\ i \ne j}}^{N}(m_i-m_j)}.
\eea 

Using matrix determinant lemma \eqref{eq:matrix_det_lemma}, the determinant of $\tilde{\gamma}_{ij}$ will take a simple form
\be
	\label{eq:det_gammaxx}
	\det(\tilde{\gamma}_{ij}x_ix_j)
	=\frac{1}{b\, A(\lambda)}\prod_{i=1}^{N}x_i^2
\ee
Finally inserting \eqref{eq:prod_h} and \eqref{eq:det_gammaxx} in \eqref{eq:det_g_initial}, we find
\be
	\label{eq:det_g}
	-\det{g}=c^{\prime}\left(\prod_{a=1}^{N-1}\lambda_a\right)^2\left(\prod\limits_{a<b}^{N-1}(\lambda_b-\lambda_{a})\right)^2\,,
\ee
with
\be
\qquad c^{\prime}=4^{1-N}\left(b\prod\limits_{i < j}^{N}(m_i-m_j)\right)^{-2}
\ee
Equipped with \eqref{eq:nhemp_inv} and \eqref{eq:det_g}, we can rewrite the Klein-Gordon equation \eqref{eq:KG}, and noting two important relations
\be\label{Key-relations-1}
	\pdiff_{\lambda_{c}}\sqrt{-\det{g}}=\left(\frac{1}{\lambda_{c}}+\sum_{\substack{b=1\\b\ne c}}^{N-1}\frac{1}{\lambda_{c}-\lambda_{b}}\right)\sqrt{-\det{g}}\,,
	\ee
	\be
	\pdiff_{\lambda_{a}}h^a=\left(-\frac{1}{\lambda_{a}}-\sum_{i}^N\frac{1}{m_i-\lambda_{a}}-\sum_{\substack{b=1\\b\ne a}}^{N-1}\frac{1}{\lambda_{a}-\lambda_{b}}\right)h^a\,,
\ee
we rewrite the Klein-Gordon equation on NHEMP metric 
\bea
    &&+\frac{1}{A(\lambda)}\left(-\frac{1}{r^2}\left[\frac{\pdiff}{\pdiff\tau}-\sum_{i=1}^{N}r k^i\frac{\pdiff}{\pdiff\varphi^i}\right]^2\varPhi+r^2\pdiff_r^2\varPhi+2r\pdiff_r\varPhi\right)\nonumber\\
	&&+\sum_{a=1}^{N-1}h^a\pdiff^2_{\lambda_{a}}\varPhi
	-\sum_{a=1}^{N-1}\sum_{i=1}^N\frac{h^a}{m_i-\lambda_{a}}\pdiff_{\lambda_{a}}\varPhi\nonumber\\
	&&+\sum_{i=1}^N\frac{1}{x_i^2}\pdiff^2_{\varphi_i}\varPhi
	-\sum_{i,j=1}^N\frac{\sqrt{m_i-1}}{m_i}\frac{\sqrt{m_j-1}}{m_j}\pdiff_{\varphi_i}
	\pdiff_{\varphi_j}\varPhi=M^2\varPhi\,.
\nonumber\\
\eea
To separate the variables in this equation, we apply the following ansatz
\be
	\label{eq:ansatz2}
	\varPhi=R_r(r) \cdot \prod_{a=1}^{N-1}R_{\lambda_{a}}(\lambda_{a})\cdot e^{i \omega \tau} \cdot \prod_{b=1}^{N}e^{i L_{b}\varphi_b}\,,
\ee
where $\omega$ and $L_i$ are arbitrary constants. In this form $\lambda_{a}$ and $r$ derivatives of the scalar $\varPhi$ will be
\bea
    &&\pdiff_{r}\varPhi=\frac{R^{'}_{r}}{R_{r}}\varPhi,
	\qquad
	\pdiff^2_{r}\varPhi=\frac{R^{''}_{r}}{R_{r}}\varPhi\,,\nonumber\\
	&&\pdiff_{\lambda_{a}}\varPhi=\frac{R^{'}_{\lambda_{a}}}{R_{\lambda_{a}}}\varPhi\,,
	\qquad
	\pdiff^2_{\lambda_{a}}\varPhi=\frac{R^{''}_{\lambda_{a}}}{R_{\lambda_{a}}}\varPhi\,,
\eea
and the Klein-Gordon equation transforms into
\bea
	\left[\vphantom{\sum_{i,j}\frac{\sqrt{m_i-1}}{m_i}}\right.
	&&\frac{1}{A(\lambda)}\left(r^2\frac{R^{''}_{r}}{R_{r}}+2r\frac{R^{'}_{r}}{R_{r}}+\frac{1}{r^2}\,(\omega-r\sum_{i=1}^N k^i L_i)^2\right)\nonumber\\
	&&+\sum_{a=1}^{N-1}h^a  \frac{R^{''}_{\lambda_{a}}}{R_{\lambda_{a}}}
	-\sum_{a=1}^{N-1}\sum_{i=1}^N\frac{h^a}{m_i-\lambda_{a}} \frac{R^{'}_{\lambda_{a}}}{R_{\lambda_{a}}}\\
	&&\left.-\sum_{i=1}^N\frac{L^2_{i}}{x_i^2}
	+\sum_{i,j=1}^N\frac{\sqrt{m_i-1}}{m_i}\frac{\sqrt{m_j-1}}{m_j}L_{i}L_{j}
	\right]\varPhi
	=M^2\varPhi\,.\nonumber\\
\eea
One can see that the first term in the above equation only depends on $r$. The separability requirement forces it to be a constant, 
\be
	\label{eq:radial}
	r^2\frac{R^{''}_{r}}{R_{r}}+2r\frac{R^{'}_{r}}{R_{r}}+\frac{1}{r^2}\,\left(\omega-r\sum_{i=1}^N k^i L_i\right)^2=\mathcal{C}_2\,.
\ee
To write the radial equation \eqref{eq:radial} as a known differential equation, we change the radial variable $r$ to $z$ by
\be
z=\frac{2\,i\,\omega}{r}\,,
\ee
which brings the equation \eqref{eq:radial} into the familiar form of Whittaker's equation 
\be\label{eq:Whittaker}
\frac{d^2 R_r}{dz^2}+\left(-\frac14+\frac{K}{z}+\frac{(1/4-\mu^2)}{z^2}\right)\,R_r=0\,,
\ee
with 
\be\label{mu-K}
K=i \sum_{j=1}^N k^j L_j\,, \quad \text{and} \quad \mu^2=\frac{1}{4}-\left(\sum_{i=1}^N k^i L_i\right)^2+\mathcal{C}_2\,.
\ee
The general solutions to the equation \eqref{eq:Whittaker} are Whittaker's functions : $\mathcal{M}_{K,\mu}(z)$ and $\mathcal{W}_{K,\mu}(z)$. These are related the confluent hypergeometric functions.%
It is interesting to study the behaviour of these functions in small and large $z$ region. (we keep $K,\mu$ parameter fixed and generic.)
For the small $z$ region, which is related to $r\to \infty$ region in the space-time, we have
\bea
&&\mathcal{M}_{K,\mu}(z)\sim z^{\mu+1/2}\,,\quad \text{as $z\to 0$}\nonumber\\
&&\mathcal{W}_{K,\mu}(z)\sim \alpha_1\,z^{\mu+1/2}+\alpha_2\, z^{-\mu+1/2}\,,\text{\footnotesize{   if $ Re(\mu)<\frac12$}}\nonumber\\
&&\mathcal{W}_{K,\mu}(z)\sim \, z^{-\mu+1/2}\text{\footnotesize{   if $Re(\mu)\ge\frac12$}}
\eea
here, $\alpha_{1,2}$ are some constants related to $\mu,K$. The asymptotic behaviour ($z \to \infty$) of $\mathcal{M}_{K,\mu}(z)$ and $\mathcal{W}_{K,\mu}(z)$ which is corresponding to $r\to 0$ region of space-time is given as 
\bea
&&\mathcal{M}_{K,\mu}(z)\sim z^{-K}e^{z/2}\nonumber\\
&&\mathcal{W}_{K,\mu}(z)\sim z^{K}e^{-z/2}\,,\quad \text{as $z\to \infty$}
\eea

Depending on the values of $\mu$ and $K$ in \eqref{mu-K}, the behaviour of the solution is different : $\mathcal{M}_{K,\mu}$ is blowing up in the $r\to 0$ ($z\to \infty$) limit, so we discard it. In the small z, i.e. ($r\to \infty$), $\mathcal{W}_{K,\mu}$ is vanishing if $Re(\mu)<1/2$ which is guaranteed by e.g. $\mathcal{C}_2<0$.

The rest of the variables can be separated after applying the relations \eqref{eq:relation_1}, \eqref{eq:relation_2} and \eqref{eq:relation_3} to the Klein-Gordon equation which will be transformed to 
\be
	\label{eq:kg_final}
	\sum_{a=1}^{N-1}\frac{P_{\lambda_{a}}(\lambda_{a})-M^2(-\lambda_a)^{N-2}}{\prod_{b=1;a\ne b}^{N-1}(\lambda_b-\lambda_a)}=0\,,
\ee
where $P_{\lambda_{a}}(\lambda_{a})$ are defined by
\be
	\begin{aligned}
	\hspace{-2mm}P_{\lambda_{a}}(\lambda_{a})\equiv&
	-\frac{4}{\lambda_{a}}\left(\frac{R^{''}_{\lambda_{a}}}{R_{\lambda_{a}}}
	-\frac{R^{'}_{\lambda_{a}}}{R_{\lambda_{a}}}\sum_{i}^N\frac{1}{m_i-\lambda_{a}}\right)\prod_{j=1}^{N} (m_j-\lambda_a)\\
	&+\frac{b}{\lambda_a}\mathcal{C}_2\prod\limits_{i=1}^{N}m_i
	+(-1)^{N-1}\sum_{i=1}^N\frac{g_{\varphi_i}}{m_i-\lambda_a}\\
	&+g_0(-\lambda_a)^{N-2}\,.
	\end{aligned}
\ee
here $g_{\varphi_i}$ and $g_0$ are some constants, defined by
\be
	g_{\varphi_i}\equiv \frac{k^2_{\varphi_i}}{m_i}\prod_{\substack{j=1\\  j\neq i}}^{N}(m_i-m_j)\,,\qquad
	g_0\equiv \left(\sum_{i=1}^{N}\frac{\sqrt{m_i-1}}{m_i}L_{i}\right)^2\,.
\ee
Equation \eqref{eq:kg_final} is only satisfied when
\be
	P_{\lambda_{a}}(\lambda_{a})=\sum_{\alpha=1}^{N-1}k_\alpha\lambda_a^{\alpha-1},
\ee
where $k_\alpha\ (\alpha=1,...,N-2)$ are arbitrary constants and $k_{N-1}=(-1)^{N-2}M^2$.

\section{Klein-Gordon equation on near horizon EVH geometry}
The metric of near horizon EVH Myers-Perry in the elliptical coordinates parameterized with $(\tau,\rho,\psi,\lambda_a,\varphi_i)$, in $d=2N+1$ dimensions is of the form
\bea\label{NHEVH-MP}
ds^2=&&F(\lambda)\left(-\rho^2d\tau^2+\frac{d\rho^2}{\rho^2}+\rho^2d\psi^2\right)\nonumber\\
    &&+\sum_{a=1}^{N-1} \hat{h}_a d\lambda_a^2
    +\sum_{a,b=1}^{N-1}\hat{\gamma}_{ab}\, \hat{x}_a(\lambda) \hat{x}_b(\lambda) d\varphi_ad\varphi_b\,,\nonumber\\
\eea
where the metric functions are

\bea
&&F(\lambda)=\prod_{a=1}^{N-1}\frac{\lambda_a}{m_a},
\qquad 
\hat{h}_a=\frac{1}{4}
\frac{{\prod_{{b \neq a}}^{N-1}}\left(\lambda_b-\lambda_a\right)}
    {\prod_{c=1}^{N-1}\left(m_c-\lambda_a\right)}\,,\nonumber\\
&&\hspace{1cm}\hat{\gamma}_{ab}=\delta_{ab}+ \frac{1}{F(\lambda)}\frac{\hat{x}_a(\lambda)}{\sqrt{m_a}}  \frac{\hat{x}_b(\lambda)}{\sqrt{m_b}}\,,
\eea
with
\be
\begin{aligned} \label{x-EVH}
\hat{x}_a^2(\lambda)=\frac{\prod\limits_{b=1}^{N-1}(m_a-\lambda_b)}{\prod\limits_{l\neq a}^{N-1}(m_a-m_l)} \,,\qquad \sum_{a=1}^{N-1}\frac{1}{m_a}=1.
\end{aligned}
\ee
(We start this section in elliptical coordinates to avoid writing multiple metrics. For more details, we refer the reader to \cite{DeBoer2012,Hakobyan:2017qee,Demirchian2018}.)

The inverse of metric \eqref{NHEVH-MP} is 
\bea
\left(\frac{\partial}{\partial s}\right)^2=&&
\frac{1}{F(\lambda)}\left(-\frac{1}{\rho^2}(\frac{\partial}{\partial \tau})^2+\rho^2(\frac{\partial}{\partial \rho})^2+\frac{1}{\rho^2}(\frac{\partial}{\partial \psi})^2\right)\nonumber\\
&&+\sum_{a=1}^{N-1}{\hat{h}}^a(\frac{\partial}{\partial \lambda_a})^2
+\sum_{a,b=1}^{N-1}\frac{\hat{\gamma}^{ab}}{\hat{x}_a \hat{x}_b}\frac{\partial}{\partial \varphi^a}\frac{\partial}{\partial \varphi^b}\,,
\eea
where
\be
\hat{\gamma}^{ab}=\delta_{ab} - \frac{\hat{x}_a(\lambda)}{\sqrt{m_a}}  \frac{\hat{x}_b(\lambda)}{\sqrt{m_b}} 
,
\qquad 
\hat{h}^a=\left(\hat{h}_a\right)^{-1}\,.
\ee
To study the Klein-Gordon equation on this metric, 
\be
\square\Phi=\frac{1}{\sqrt{-g}}{\pdiff_\alpha(\sqrt{-g}g^{\alpha\beta}\pdiff_\beta\Phi)}=M^2\Phi\,,
\ee
we need to compute the determinant of the metric,
\be
-\det{g}=\rho^2F_0(\lambda)^3\left(\prod_{a=1}^{N-1}\hat{h}_a\right)\det{(\hat{\gamma}_{ab}\hat{x}_a\hat{x}_b)}\,,
\ee
It is easy to show that 
\be
\prod_{l=1}^{N-1}\hat{h}_l=\hat{c}\cdot\frac{\prod\limits_{\substack{a,b=1\\a\ne b}}^{N-1}\left(\lambda_b-\lambda_a\right)}{\prod\limits_{a=1}^{N-1}\hat{x}_a^2}, \quad \hat{c}=4^{1-N}\prod_{\substack{a,b=1\\a\ne b}}^{N-1}\left(m_a-m_b\right)^{-1}\,,
\ee

Using matrix determinant lemma \eqref{eq:matrix_det_lemma}, we have
\be
\det{(\hat{\gamma}_{ab}\hat{x}_a\hat{x}_b)}=\frac{1}{F(\lambda)}\prod_{a=1}^{N-1}\hat{x}_a^2\,.
\ee
Finally, the determinant of metric \eqref{NHEVH-MP} simplifies to
\be
-\det{g}=c''\,\rho^2\,\left(\prod_{l=1}^{N-1}\lambda_l\right)^2\left(\prod_{\substack{a,b=1\\b<a}}^{N-1}(\lambda_b-\lambda_a)\right)^2\,,
\ee
with
\be
c''=\frac{1}{4^{N-1}}\left(\prod_{l=1}^{N-1}{m_l}\right)^{-2}\left(\prod_{\substack{a,b=1\\b<a}}^{N-1}(m_b-m_a)\right)^{-2}\,.
\ee
The key relations for the separability of the Klein-Gordon equation are as follows

\be\label{Key-relations-2}
\begin{aligned}
&	\pdiff_{\lambda_{a}}\sqrt{-\det{g}}=\left(\frac{1}{\lambda_{a}}-\sum_{\substack{b=1\\b\ne a}}^{N-1}\frac{1}{\lambda_{b}-\lambda_{a}}\right)\sqrt{-\det{g}}\,,\\
&   \pdiff_{\lambda_{a}}\hat{h}^a=\left(-\sum_{b=1}^{N-1}\frac{1}{m_b-\lambda_{a}}+\sum_{\substack{b=1\\b\ne a}}^{N-1}\frac{1}{\lambda_{b}-\lambda_{a}}\right)\hat{h}^a\,.
\end{aligned}
\ee

The Klein-Gordon equation becomes
\bea
\label{expanded-KG}
	&&\frac{1}{F(\lambda)}\left(-\frac{\partial_\tau^2\varPhi-\partial_\psi^2\varPhi}{\rho^2}+\rho^2\partial_\rho^2\varPhi+3\rho\partial_\rho\varPhi\right)\nonumber\\
	&&+\sum_{a=1}^{N-1}\frac{1}{\hat{x}_a^2}\pdiff^2_{\varphi_a}\varPhi-\sum_{a,b}^{N-1}\frac{1}{\sqrt{m_a}}\frac{1}{\sqrt{m_b}}\pdiff_{\varphi_a}\pdiff_{\varphi_b}\varPhi\nonumber\\
	&&+\sum_{a=1}^{N-1}\hat{h}^a\pdiff^2_{\lambda_{a}}\varPhi
	+\sum_{a=1}^{N-1}\frac{\hat{h}^a}{\lambda_{a}}\pdiff_{\lambda_{a}}\varPhi
	-\sum_{a,b=1}^{N-1}\frac{\hat{h}^a}{m_b-\lambda_{a}}\pdiff_{\lambda_{a}}\varPhi=M^2\varPhi\,.\nonumber\\
\eea
here $\hat{x}^a$ is defined in \eqref{x-EVH}.
To separate the variables, we use the following ansatz for the scalar field
\be
	\label{eq:ansatz}
	\varPhi=R_\rho(\rho) \cdot \prod_{a=1}^{N-1}R_{a}(\lambda_{a})\cdot e^{i (-k_\tau \tau+m_\psi\psi)} \cdot \prod_{b=1}^{N-1}e^{i L_{b}\varphi_b}\,,
\ee
into the Klein-Gordon equation \eqref{expanded-KG}. We observe that the first parentheses includes only $\rho$-dependent term. The separability requires it to be a constant 

\be\label{eq:radial-rho}
\left(\frac{k_\tau^2-m_\psi^2}{\rho^2}+\rho^2\partial_\rho^2+3\,\rho\,\partial_\rho\right)R_\rho(\rho)=-4\,\hat{\mathcal{C}}_2\,R_\rho(\rho)\,.
\ee
Replacing 
\bea\label{change-rho}
R_\rho(\rho)=\frac{u(\rho)}{\rho}\,,\quad \text{and}\quad \rho=\frac{\sqrt{k_\tau^2-m_\psi^2}}{z}\,,
\eea
into the radial equation \eqref{eq:radial-rho}, we get the Bessel's equation
\bea
z^2\frac{d^2}{dz^2}u+z\frac{d}{dz}u+\left(z^2-\nu^2\right)u=0
\eea
with
\bea\label{nu-parameter}
\nu^2=1-4\hat{\mathcal{C}}_2
\eea
whose solutions are Bessel functions $J_\nu(z)$, $Y_\nu(z)$.
The small and large $r$ behaviour of the solution is dictated by the asymptotic behaviour of Bessel functions. In the small $r$ region of space-time which is related to $z\to\infty$, they give
\bea
&&J_\nu(z)\sim \frac{1}{\sqrt{z}}cos(z-\frac{\pi \nu}{2}-\frac{\pi}{4})\,,\nonumber\\
&&Y_\nu(z)\sim \frac{1}{\sqrt{z}}sin(z-\frac{\pi \nu}{2}-\frac{\pi}{4})\,, \text{as $z\to\infty$}\,,
\eea
which is vanishing. Also, in the large $r$ region of space-time, corresponding to $z\to 0$, they behave as
\bea
&&J_\nu(z)\sim \frac{1}{\Gamma(\nu+1)}\left(\frac{z}{2}\right)^\nu\,,\nonumber\\
&&Y_{\nu}(z) \sim -\frac{\Gamma(\nu)}{\pi}\left(\frac{z}{2}\right)^{-\nu}\,.
\eea
In the case of $\hat{\mathcal{C}}_2<0$ which leads to $\nu>1$, $Y_\nu$ blows up while $J_\nu$ falls off in the large $r$ region and is an acceptable solution.

For the rest of the Klein-Gordon equation, we have
\bea
	\left(\vphantom{\sum_{i,j}\frac{\sqrt{m_i-1}}{m_i}}\right.
	&&-\frac{4\hat{\mathcal{C}}_2}{F_0(\lambda)}-\sum_{a,b=1}^{N-1}\frac{L_{a}}{\sqrt{m_a}}\frac{L_{b}}{\sqrt{m_b}}\nonumber\\
	&&\left.+\sum_{a=1}^{N-1}\left[\hat{h}^a  \frac{R^{''}_{a}}{R_{a}}
	+\frac{\hat{h}^a}{\lambda_{a}} \frac{R^{'}_{a}}{R_{a}} 	
	-\sum_{i=1}^{N-1}\frac{\hat{h}^a}{m_i-\lambda_{a}} \frac{R^{'}_{a}}{R_{a}}
	-\frac{k^2_{\varphi_a}}{x_a^2}\right] \right)\varPhi\nonumber\\
	&&\hspace{7.5cm}=M^2\varPhi\,.\nonumber\\
\eea
Using the relations \eqref{eq:relation_1}, \eqref{eq:relation_2} and \eqref{eq:relation_3}, we see that the Klein-Gordon equation is separable
\be
	\label{eq:kg_final-EVH}
	\sum_{a=1}^{N-1}\frac{\hat{Q}_{\lambda_a}(\lambda_{a})-M^2(-\lambda_a)^{N-2}}{\prod_{b=1;a\ne b}^{N-1}(\lambda_b-\lambda_a)}=0\,,
\ee
where 
\be
	\begin{aligned}
	\hspace{-4mm}\hat{Q}_{\lambda_a}(\lambda_{a})=&\,
 	 4\left(\frac{R^{''}_{a}}{R_{a}}+ \frac{1}{\lambda_a}\frac{R^{'}_{a}}{R_{a}}\
	-\sum_{b=1}^{N-1}\frac{R^{'}_{a}/R_{a}}{m_b-\lambda_{a}}\right)\prod_{c=1}^{N-1} (m_c-\lambda_a)\\
	&-\frac{4\hat{\mathcal{C}}_2}{\lambda_a}\prod\limits_{b=1}^{N-1}m_b+\sum_{b=1}^{N-1}\frac{\hat{q}_{\varphi_b}}{m_b-\lambda_a}
	+\hat{q}_0(-\lambda_a)^{N-2}\,,
	\end{aligned}
\ee
and
\be
	\hat{q}_{\varphi_a}\equiv(-1)^{N-1}\,k^2_{\varphi_a}\prod_{\substack{b=1\\ b\neq a}}^{N-1}(m_a-m_b)\,,\quad
	\hat{q}_0\equiv\left(\sum_{a=1}^{N-1}\frac{L_{a}}{\sqrt{m_a}}\right)^2\ .
\ee
Equation \eqref{eq:kg_final-EVH} is only satisfied when
\be
	\hat{Q}_{\lambda_{a}}(\lambda_{a})=\sum_{\alpha=1}^{N-1}k_\alpha\lambda_a^{\alpha-1},
\ee
where $k_\alpha\ (\alpha=1,...,N-2)$ are arbitrary constants and $k_{N-1}=(-1)^{N-2}M^2$.

\section{Discussions}
We studied the separability of Klein-Gordon equation on two near horizon geometries in d-dimensions : generic extremal and extremal vanishing horizon cases. Since the latter case exists only in odd dimensions, we take $d=2n+1$ in both cases. We do not expect that the separability of Klein-Gordon equations changes for the first case if we take even-dimensions. This expectation roots from the fact that the separability of geodesic equation has been shown for both even and odd dimensions \cite{Demirchian2018} and the Klein-gordon and geodesic equation are related in the semi-classical limit : if we write the solution to the Klein-Gordon equation $\Phi$ as $\Phi=\mathcal{N}\, exp\left(\frac{i S}{\alpha}\right)$, we get the Hamilton-Jacobi equation for $S$ in the so-called semi-classical limit, $\alpha \to 0$.
(see \cite{Sergyeyev:2007gf} for an explicit example.)

It is worth mentioning that since the extremal/EVH limit does not commute with near horizon limit, we \emph{could not} get the Klein-Gordon equation on near horizon EVH geometry by taking the EVH limit on NHEMP metric. Therefore, we studied the Klein-Gordon equation on near horizon EVH geometry, independently.

One of the interesting problems to which we will come back in future is to study the cases when some or all of the rotation parameters are equal.

Regarding the solution to the radial equations, we discussed the asymptotic behaviour for the generic values of the parameters. There are some special limits in the parameter space which can lead to different asymptotic behaviour (like $\mu=0$ of $\mathcal{M}_{K,\mu}$ or $\mathcal{W}_{K,\mu}$). These special limits are known for Whittaker and Bessel functions. We are interested in the interpretation of their consequences in near horizon geometries.

Our next step is to study the other probes on these near horizon geometries such as Dirac, Maxwell and p-form fields and show the separability of their field equations.


\section*{Acknowledgement}
We would like to thank M.M. Sheikh-Jabbari and A. Nersessian for the fruitful discussions during our previous collaboration, their contribution at the first stages of this project and comments on the draft. H.D. acknowledges the support of ICTP program Network scheme NT-04 for his visit to School of Physics, IPM.

\appendix

\section{Useful Identities}\label{app-A}
The following identity holds between $N-1$ independent variables $\lambda_a$ and any parameter $\kappa$
\be
\begin{aligned}
	\label{eq:relation}
	\sum_{a=1}^{N-1}\frac{1}{\prod\limits_{\substack{b=1\\{b\ne a}}}^{N-1}(\lambda_a-\lambda_b)}\frac{(\lambda_a)^\alpha}{\kappa-\lambda_a}=\frac{(\kappa)^\alpha}{\prod\limits_{a=1}^{N-1}(\kappa-\lambda_a)}-\delta_{\alpha,N-1}\,,
\end{aligned}
\ee
where $\alpha=\{0,...,N-1\}$. For $\alpha = 0$, this equation reduces to
\be
    \label{eq:relation_1}
	\frac{1}{\prod\limits_{a=1}^{N-1}(\lambda_a-\kappa)} =   \sum_{a=1}^{N-1}\frac{1}{\prod\limits_{b=1;a\ne b}^{N-1}(\lambda_b-\lambda_a)}\frac{1}{\lambda_a-\kappa}    
\ee
and for an additional condition of $\kappa = 0$ we get
\be
    \label{eq:relation_2}
    \frac{1}{\lambda_1\ldots\lambda_{N-1}}= \sum_{a=1}^{N-1} \frac{1}{\prod_{b=1; b\neq a}^{N-1}(\lambda_b-\lambda_a)}\frac{1}{\lambda_a}  
\ee
On the other hand, setting $\kappa = 0$ in \eqref{eq:relation} results into
\be
	\label{eq:relation_3}
	\sum_{i=1}^{N} \frac{\lambda^\beta_i}{\prod_{j=1; i\neq j}^{N}(\lambda_i-\lambda_j)}=\delta_{\beta,
		N-1}
	\quad
	{\rm for  }\quad 0\leq  \beta\le N-1.
\ee
Another relation that we use in this paper is the so called matrix determinant lemma which states
\be
\label{eq:matrix_det_lemma}
    \det(\mathrm{I}+\boldsymbol{xy}^T)=1+\boldsymbol{y}^T\boldsymbol{x}
\ee
where $\mathrm{I}$ is a unit matrix and $\boldsymbol{xy}^T$ is the outer product of two vectors $\boldsymbol{x}$ and $\boldsymbol{y}$.
\bibliographystyle{fullsort.bst}

\end{document}